\documentclass[%
 aip,
 amsmath,amssymb,
 reprint,%
]{revtex4-1}

\usepackage{graphicx}
\usepackage{dcolumn}
\usepackage{bm}

\usepackage[utf8]{inputenc}
\usepackage[T1]{fontenc}
\usepackage{mathptmx}
\usepackage{etoolbox}

\usepackage{xcolor}

\makeatletter
\def\@email#1#2{%
 \endgroup
 \patchcmd{\titleblock@produce}
  {\frontmatter@RRAPformat}
  {\frontmatter@RRAPformat{\produce@RRAP{*#1\href{mailto:#2}{#2}}}\frontmatter@RRAPformat}
  {}{}
}%
\makeatother
\begin{document}


\title[Network modelling of yield-stress fluid flow in randomly disordered porous media]{Network modelling of yield-stress fluid flow in randomly disordered porous media}

\author{Cl\'{a}udio P.~Fonte}
\affiliation{Department of Chemical Engineering, The University of Manchester, Oxford Road, Manchester M13~9PL, UK}

\author{Elliott Sutton}
\affiliation{Department of Chemical Engineering, The University of Manchester, Oxford Road, Manchester M13~9PL, UK}
\affiliation{Manchester Centre for Nonlinear Dynamics, The University of Manchester, Oxford Road, Manchester M13~9PL, UK}

\author{Kohei Ohie}
\affiliation{Laboratory for Flow Control, Division of Mechanical and Aerospace Engineering, Faculty of Engineering, Hokkaido University, Sapporo N13W8, Hokkaido 060-8628, Japan}

\author{Eleanor Doman}
\affiliation{Department of Mathematics, The University of Manchester, Oxford Road, Manchester M13~9PL, UK}

\author{Yuji Tasaka}
\affiliation{Laboratory for Flow Control, Division of Mechanical and Aerospace Engineering, Faculty of Engineering, Hokkaido University, Sapporo N13W8, Hokkaido 060-8628, Japan}

\author{Anne Juel}
\email{anne.juel@manchester.ac.uk}
\affiliation{Manchester Centre for Nonlinear Dynamics, The University of Manchester, Oxford Road, Manchester M13~9PL, UK}
\affiliation{Department of Physics and Astronomy, The University of Manchester, Oxford Road, Manchester M13~9PL, UK}

\date{\today}

\begin{abstract}
Yield-stress fluid flow through porous media is governed by a strong coupling between rheology and pore-scale geometry, leading to nonlinear, non-Darcy transport and pronounced channelisation near yielding. We develop a pore-network model for Herschel-Bulkley flow in two-dimensional disordered porous media, including optional wall slip. The network is closed by a physics-based pressure-flow relation for a converging-diverging throat, so that yielding and post-yield transport emerge directly from the pore-scale fluid mechanics without fitted resistance parameters. Benchmarking against direct numerical simulations shows that the model captures both the bulk pressure drop and the evolution of the flow topology from spatially distributed transport to strongly channelised flow. The framework also captures the leading effect of wall slip, which lowers the pressure gradient required for transport and reactivates pathways that remain blocked in the no-slip case. Using the model across different porous geometries, we show that near-yield pressure losses are governed by constriction statistics rather than by an obstacle-scale length. In particular, rescaling with the domain-averaged minimum throat width collapses the plastic-dominated response across porosities, identifying the dissipation-relevant geometric scale for viscoplastic transport in this regime.
\end{abstract}

\maketitle

The flow of yield-stress fluids through porous media is central to applications including enhanced oil recovery~\citep{Prud-homme:1996,Zhou-et-al:2019}, filtration \citep{Meeten:1994,Sochi:2010}, coating penetration~\citep{Germani-et-al:2007,Mori-et-al:2017}, and soil remediation~\citep{Oliveira-et-al:2004,Forey-et-al:2020,Castro-et-al:2023}. In these systems, non-Darcy behaviour is intrinsic: flow begins only above a finite pressure threshold, the relation between flow rate and pressure drop is nonlinear, and transport near yielding becomes highly heterogeneous because only part of the pore space is mobilised \citep{talon-and-bauer:2013,Chaparian-Tammisola:2021,Schimmenti-et-al:2023}. Wall slip, commonly observed in soft glassy and particulate materials, can add further complexity to this behaviour by reducing wall friction, lowering the apparent yield threshold, and suppressing channelisation when slip is sufficiently strong \citep{Zhang2017,Meeker-Bonnecaze-Cloitre:2004a,Seth-Cloitre-Bonnecaze:2008,Walls-et-al:2003,Chaparian-Tammisola:2021}. \textcolor{black}{In these materials, slip is generally interpreted as an apparent macroscopic boundary condition at solid surfaces, caused by near-wall microstructural mechanisms such as particle or microgel depletion, solvent-rich lubrication layers, and rearrangement or deformation of soft particles adjacent to smooth boundaries.} Predicting how yielding, pore-scale disorder, and slip combine to determine the macroscopic flow response remains a major challenge.

Direct numerical flow simulations can resolve these effects, but their cost and numerical complexity limit their use in large or strongly disordered porous domains. Pore-network models offer an efficient reduced-order description by representing the void space as pores connected by throats governed by local pressure-flow relations \citep{Fatt:1956,Hunt-Ewing:2009}. Their predictive power therefore depends critically on the fidelity of the throat-scale closure. This is especially true near the yielding transition, where the global response is controlled by a small subset of constrictions and even modest local modelling errors can lead to large errors in the predicted threshold and post-yield transport. Existing network models for viscoplastic flow capture some qualitative trends \citep{Balhoff-and-Thompson:2004,Sochi:2010}, but discrepancies remain pronounced close to yielding \citep{chase-and-dachavijit:2003,Park:1972,Al-Fariss-Pinder:1984}, and some formulations rely on fitted resistance parameters \citep{waisbord-et-al:2019,Liu-et-al:2019}. A predictive network model that incorporates both yield-stress rheology and wall slip without empirical calibration is, to the best of our knowledge, still lacking in the literature.

In this Letter, we develop a pore-network model for Herschel-Bulkley flow in two-dimensional disordered porous media. The network is closed by a physics-based pressure-flow relation for a converging-diverging throat, with optional wall slip, so that both yielding and post-yield transport emerge directly from the pore-scale fluid mechanics without fitted resistance parameters. Comparison against direct numerical simulations shows that the model captures the macroscopic departure from Darcy behaviour through the yielding transition. The framework further reveals that near-yield pressure losses are controlled by constriction statistics rather than by an obstacle-scale length: rescaling with the domain-averaged minimum throat width collapses the plastic-dominated response across porosities, identifying the dissipation-relevant geometric scale in this regime.

\begin{figure}[h]
    \centering
    \includegraphics[width=\linewidth]{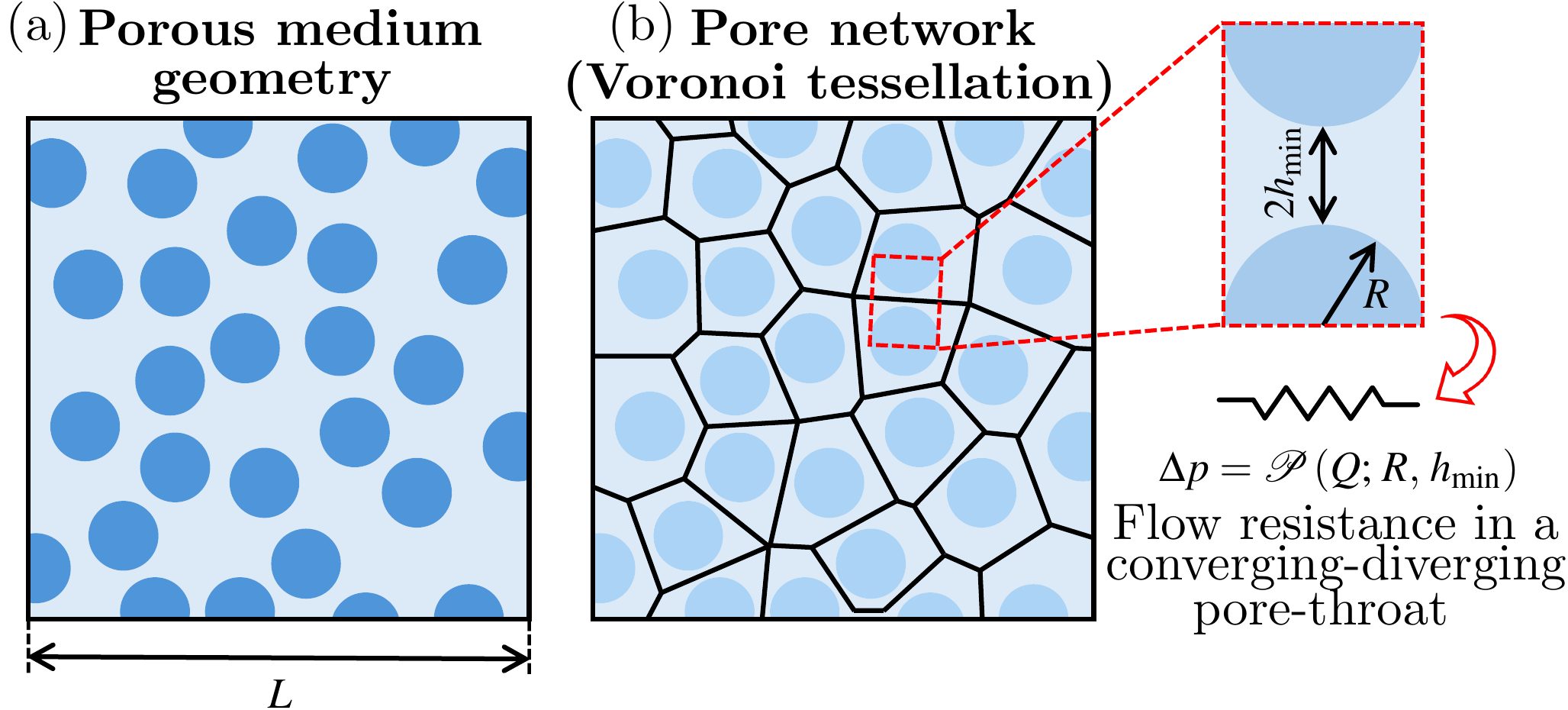}
    \caption{(a) Two-dimensional porous medium formed by randomly placed circular obstacles. (b) Voronoi-based pore network: vertices are pores and edges are throats. \textcolor{black}{The highlighted throat is defined by a neighbouring pair of obstacles; their centre-to-centre distance is $2(h_{\mathrm{min}}+R)$, where $h_{\mathrm{min}}$ is the minimum throat half-width and $R$ is the obstacle radius. 
    The throat resistance is represented by the pressure-flow relation $\Delta p=\mathcal{P}(Q)$ for the corresponding converging-diverging gap [see \eqref{eqn:pressure_drop_function}].}}
    \label{fig:geometrynetwork}
\end{figure}
\begin{table}[htbp]
\caption{\label{tab:geometric parameters} Geometric characteristics of the porous media considered. Case CT is identical to that studied by \citet{Chaparian-Tammisola:2021}.}
\begin{ruledtabular}
\begin{tabular}{ccccc}
Case & $L/R$ & $\phi$ & $\bar{h}_\mathrm{min}/R$  & \textcolor{black}{$\sigma_{h_\mathrm{min}}/R$}\\
     \hline 
CT & 50 & 0.7 & 0.795 & \textcolor{black}{0.491}\\
G1 & 20 & 0.600 & 0.564 & \textcolor{black}{0.355}\\
G2 & 20 & 0.702 & 0.830 & \textcolor{black}{0.513}\\
G3 & 20 & 0.804 & 1.23 & \textcolor{black}{0.756}\\
\end{tabular}
\end{ruledtabular}
\end{table} 

We start by considering two-dimensional porous domains of size $L \times L$ formed by the void space between randomly placed, non-overlapping circular obstacles of radius $R$ [see Fig.~\ref{fig:geometrynetwork}(a)]. Table~\ref{tab:geometric parameters} lists the distinct realisations considered in this study, spanning a range of morphologies and porosities. \textcolor{black}{The CT case is used for direct validation against the geometry of \citet{Chaparian-Tammisola:2021}, whereas G1--G3 are smaller domains used for the parametric study of porosity, constriction statistics, and wall slip.} In addition to the obstacle radius, $R$, and domain size, $L$, each porous medium is characterised by its porosity, $\phi~=~1~-~\pi R^2 N/L^2$, where $N$ denotes the total number of obstacles in the domain. \textcolor{black}{We also characterise each network by the distribution of minimum pore-throat half-widths, $h_{\mathrm{min}}$, reporting its mean, $\bar{h}_{\mathrm{min}}$, and standard deviation, $\sigma_{h_{\mathrm{min}}}$ in Table~\ref{tab:geometric parameters}.} 

To construct a pore-network representation, we compute the Voronoi tessellation of the obstacle centroids [see Fig.~\ref{fig:geometrynetwork}(b)]. The Voronoi vertices define pores and the Voronoi edges define throats, yielding a graph $(\mathcal{V},\mathcal{E})$ with set of pores $\mathcal{V}$ and set of throats $\mathcal{E}$. Flow is modelled by mass conservation at each pore and a throat pressure-flow law on each edge. For pore $i\in\mathcal{V}$,
\begin{equation}
\label{eqn:mass_conservation}
\sum_{j} Q_{i\to j}=0,
\end{equation}
where the sum is over all neighbours $j$ connected to $i$ by $(i,j)\in\mathcal{E}$. Each edge is assigned a fixed orientation $i\to j$, and $Q_{i\to j}$ is taken positive in that direction. The pressure drop across a throat satisfies
\begin{equation}
\label{eqn:pressure_continuity}
p_i-p_j=\mathcal{P}\!\left(Q_{i\to j}\right),
\end{equation}
where $\mathcal{P}(\cdot)$ is the (generally nonlinear) pressure-flow relation associated with the converging-diverging gap between neighbouring obstacles.

Although network channels are represented as straight Voronoi edges, the converging-diverging throat shape set by the interstitial gap between two neighbouring circular obstacles is embedded in the closure $\mathcal{P}(Q_{i\to j})$. We estimate the throat-scale pressure drop by modelling steady, unidirectional flow of a generalised Newtonian fluid through the gap. Introducing local coordinates with $x$ along the Voronoi edge and $y$ transverse, the domain is $x\in[0,2R]$ and $y\in[-h(x),h(x)]$, with
$h(x)=h_{\min}+R-\sqrt{R^2-(x-R)^2}$,
where $2h_{\min}$ is the minimum pillar separation.

Under the unidirectional flow approximation, $\partial\! p/\partial\! y=0$, the streamwise momentum balance reduces to
\begin{equation}
\label{eqn:leading order solution 2}
-\frac{\partial \!}{\partial \! y}\!\left(\eta(\dot{\gamma})\,\frac{\partial \! u}{\partial \! y}\right)=G(x),
\qquad
G(x)=-\frac{\mathrm{d}p}{\mathrm{d}x},
\end{equation}
where $u(x,y)$ is the streamwise velocity and $\dot{\gamma}=|\partial\! u/\partial\! y|$. We use the Herschel-Bulkley law
$\eta(\dot{\gamma})=\tau_0/\dot{\gamma}+K\,\dot{\gamma}^{\,n-1}$ to describe viscoplasticity,
with yield stress $\tau_0$ and consistency and power-law indices $K$ and $n$. \textcolor{black}{Thus, the throat-scale closure retains the local variation of shear rate, and hence apparent viscosity, across each throat, although the full two-dimensional strain-rate field within pore bodies is not resolved explicitly.}  At the walls we impose a slip condition, $u(x,\pm h)=u_s(x)$, \textcolor{black}{using a general threshold-type apparent-slip law of the form}
\begin{equation}
\label{eqn:dimensional slip model}
u_s(x)=\frac{\alpha\left(\tau_w(x)-\tau_s\right)^\beta}{\sqrt{1+h'(x)^2}},
\qquad
\tau_w(x)=G(x)\,h(x),
\end{equation}
where $\alpha$ is the slip coefficient, $\beta$ a slip exponent (taken as $\beta=1$ in this work \citep{Seth-et-al:2012}), and $\tau_s$ a slip yield stress. \textcolor{black}{These parameters should be interpreted as effective quantities that depend on the fluid, wall material and roughness.} The factor $1/\sqrt{1+h'(x)^2}$ accounts for wall inclination by projecting tangential slip onto the streamwise direction.

The flow rate is related to the local pressure gradient in the pore throat by
\begin{equation}
\label{eqn:flow-rate}
Q=\int_{-h(x)}^{h(x)}u(x,y)\,\mathrm{d}y.
\end{equation}
Solving \eqref{eqn:leading order solution 2} subject to \eqref{eqn:dimensional slip model} yields the local relation between $Q$ and $G(x)$,
\begin{widetext}
\begin{equation}
\label{eqn:flow_solution}
Q=
\begin{cases}
\dfrac{2h\,\alpha\left(Gh-\tau_s\right)^{\beta}}{\sqrt{1+h'^2}}
+\dfrac{2nK}{G^{2}}\left(\dfrac{Gh-\tau_0}{K}\right)^{\frac{n+1}{n}}
\dfrac{(n+1)Gh+n\tau_0}{(n+1)(2n+1)} +\varepsilon(G),
& Gh>\tau_0, \\[6pt]
\dfrac{2h\,\alpha\left(Gh-\tau_s\right)^{\beta}}{\sqrt{1+h'^2}} +\varepsilon(G),
& \tau_s< Gh\le \tau_0, \\[6pt]
\varepsilon(G), & Gh\le \tau_s.
\end{cases}
\end{equation}
\end{widetext}
where $\varepsilon(G)=\frac{2}{3} G h^3/\mu_{\mathrm{reg}}$ is a regularisation parameter corresponding to the volumetric flow rate of a Newtonian fluid with viscosity $\mu_{\mathrm{reg}}$, introduced in \eqref{eqn:flow_solution} to improve numerical robustness. We choose $\mu_{\mathrm{reg}}$ sufficiently large that $\varepsilon(G)$ is negligible compared with the physical contributions in the yielded regimes, while preventing exactly zero flow.

Finally, the pressure-drop functional entering \eqref{eqn:pressure_continuity} is obtained by integrating the local pressure gradient along the pore throat at fixed flow rate. By symmetry under flow reversal, it is odd, and we write
\begin{equation}
    \label{eqn:pressure_drop_function}
    \mathcal{P}(Q)=\operatorname{sgn}(Q)\int_{0}^{2R} G\!\left(x;\lvert Q\rvert\right)\,\mathrm{d}x .
\end{equation}
Since $G(x;|Q|)$ cannot be obtained in closed form, we compute it by numerically inverting \eqref{eqn:flow_solution} for each $x$, and evaluate \eqref{eqn:pressure_drop_function} using numerical quadrature.

For a network with $N_v$ interior pores and $N_e$ throats, enforcing mass conservation \eqref{eqn:mass_conservation} together with the throat pressure-flow relation \eqref{eqn:pressure_continuity} yields a coupled system of $N_v+N_e$ nonlinear equations, written compactly as $\mathbf{F}(\mathbf{z})=\mathbf{0}$. Here $\mathbf{z}$ collects the throat flow rates together with the pressures at interior pores; inlet/outlet pore pressures are prescribed and excluded from the vector of unknowns. We write $\mathbf{F}=\big[\mathbf{f}(\mathbf{Q}),\,\mathbf{g}(\mathbf{p},\mathbf{Q})\big]^{\mathsf T}$, where the $N_v$ mass-balance residuals are
\begin{equation}
f_i=\sum_{j}Q_{i\to j}, \qquad i\in\mathcal{V},
\end{equation}
and the $N_e$ throat residuals enforce the pressure-flow relation,
\begin{equation}
g_{ij}=p_i-p_j-\mathcal{P}(Q_{i\to j}), \qquad (i,j)\in\mathcal{E}.
\end{equation}

We solve $\mathbf{F}(\mathbf{z})=\mathbf{0}$ using a Newton-type method with trust-region globalisation, implemented in Julia via \texttt{NonlinearSolve.jl}; the Jacobian is formed by central finite differences. Because the system becomes stiff near yielding, we use continuation in the imposed inlet pressure: solutions are computed along a sweep of $p_{\mathrm{in}}$, using the previous converged solution (with linear extrapolation from the last two steps) as the initial guess. Convergence is declared when the residual norm satisfies prescribed absolute and relative tolerances. The Julia implementation used in this study is openly available on GitHub~\cite{github}.

\begin{figure}[ht]
    \centering
    \includegraphics[scale=1.0]{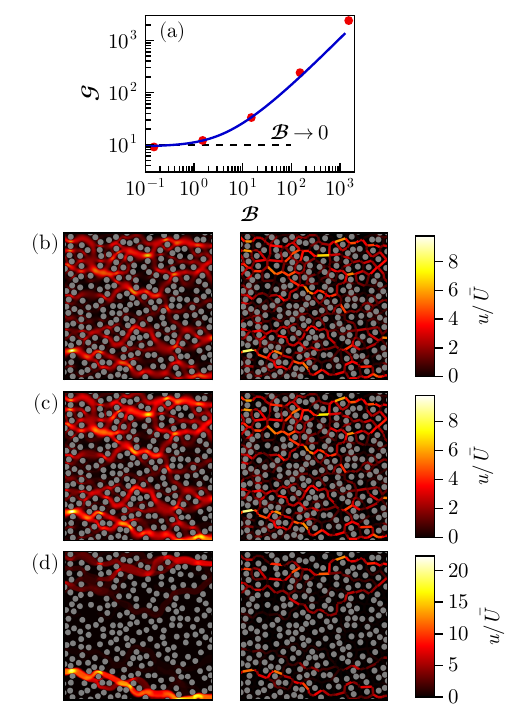}
    \caption{(a) No-slip response in geometry CT: dimensionless bulk pressure gradient $\mathcal{G}$ versus bulk Bingham number $\mathcal{B}$. Solid line, present network model; markers, direct numerical simulations of \citet{Chaparian-Tammisola:2021}. The horizontal dashed line denotes the Newtonian limit ($\mathcal{B}\to0$) from the fully resolved flow simulations. (b-d) Velocity magnitude maps in geometry CT without slip, comparing direct numerical simulations (left) and the network model (right) for (b) $\mathcal{B}=0$, (c) $\mathcal{B}=1.5$, and (d) $\mathcal{B}=150$. The left-hand panels in (c) and (d) are reproduced from \citet{Chaparian-Tammisola:2021} under CC BY 4.0.}
    \label{fig:flowrate_pressuredrop}
\end{figure}

Network-model predictions are benchmarked against two sets of fully resolved two-dimensional simulations: the viscoplastic simulations of \citet{Chaparian-Tammisola:2021}, which employ an augmented-Lagrangian finite-element formulation to accurately resolve yielded and unyielded regions, and our own direct simulations in the Newtonian limit. \textcolor{black}{These fully resolved simulations provide a stringent benchmark for the reduced-order network model, since the geometry, constitutive law, boundary conditions, and imposed pressure drop are precisely specified.} Full details of the numerical methodology for the Newtonian-limit simulations, including the discretisation and solver settings, are similar to those reported by \citet{Sutton-et-al:2022}; only the numerical results are presented here.

For consistency, we adopt the non-dimensionalisation of \citet{Chaparian-Tammisola:2021}, based on the obstacle radius $R$, and denote the bulk Bingham number and streamwise pressure gradient by $(\mathcal{B},\mathcal{G})$. Thus,
\begin{equation}
    \mathcal{G} = \frac{\Delta p}{L}\,\frac{R}{K(\bar{U}/R)^n},
    \qquad
    \mathcal{B} = \frac{\tau_0}{K(\bar{U}/R)^n},
\end{equation}
where $\Delta p$ is the imposed pressure drop across the porous domain and $\bar{U}=Q_T/L$ is the mean inlet velocity, with $Q_T$ the total flow rate percolating through the network. Unlike \citet{Chaparian-Tammisola:2021}, however, we define the mean velocity using the full domain width $L$ rather than the open inlet length $L_{\mathrm{inl}}$. Their reported values of $\mathcal{B}$ and $\mathcal{G}$ were therefore rescaled by $(L/L_{\mathrm{inl}})^n$ to match the present definition; for the Bingham-fluid data considered here, this reduces to $L/L_{\mathrm{inl}}$. \textcolor{black}{The numerical continuation becomes increasingly challenging at very large Bingham numbers. In particular, for $\mathcal{B}\gtrsim 10^3$, the nonlinear system becomes highly stiff because only a sparse percolating pathway remains active. Robustly accessing this regime would require increasingly fine continuation steps and possibly more specialised nonlinear-solver formulations, with substantially greater computational cost. Moreover, the associated flow rates become vanishingly small, so this regime lies outside the timescales typically relevant to practical transport applications. For these reasons, the present study focuses on $0 \leq \mathcal{B} \leq 10^3$.}

Fig.~\ref{fig:flowrate_pressuredrop}(a) compares the predicted no-slip response in geometry CT (cf. Table~\ref{tab:geometric parameters}) with the numerical results. The network model reproduces the monotonic increase of $\mathcal{G}$ with $\mathcal{B}$, recovering the Newtonian limit as $\mathcal{B}\to0$ and the linear viscoplastic scaling $\mathcal{G}= c\,\mathcal{B}$ at large $\mathcal{B}$, where $c$ depends on the geometry. This large-$\mathcal{B}$ regime reflects the fact that only a small number of flow pathways remain fluidised [cf. Fig.~\ref{fig:flowrate_pressuredrop}(d)]. 

Fig.~\ref{fig:flowrate_pressuredrop}(b-d) compare velocity magnitude maps over a range of $\mathcal{B}$. At small $\mathcal{B}$, the network model reproduces both the spatially widespread flow topology, with most pathways remaining active, and the distribution of local velocities. With increasing $\mathcal{B}$, it captures the onset of channelisation and the progressive shutdown of secondary pathways. Even close to the yield limit, the model retains the correct hierarchy of active pathways and reproduces their relative velocity scales well.

We find good agreement across the full range of $\mathcal{B}$ studied, particularly in view of the coarse-grained nature of the network description. The model reproduces the bulk response closely at small $\mathcal{B}$ and remains quantitatively consistent as $\mathcal{B}$ increases, although it tends to under-predict the resistance once the flow is confined to only a few active pathways. This deviation is consistent with the two main simplifications adopted here: first, the throat-scale closure is based on a reduced (one-dimensional) description of the converging-diverging gap and is least accurate for the wider constrictions that dominate transport in the strongly channelised regime; second, additional dissipation within yielded pore bodies (outside of pore-throats) is not represented explicitly. Even so, the network coupling does not amplify the local closure error, and the model retains good predictive fidelity at the macroscopic scale, with deviations of only $17\%$ at $\mathcal{B}=100$ and $31\%$ at $\mathcal{B}=10^3$. The model captures the dominant physics of yielding, channelisation, and the macroscopic departure from Darcy behaviour, providing an efficient framework for exploring these systems at a fraction of the cost of fully resolved simulations.

\begin{figure}
    \centering
    \includegraphics[scale=1.0]{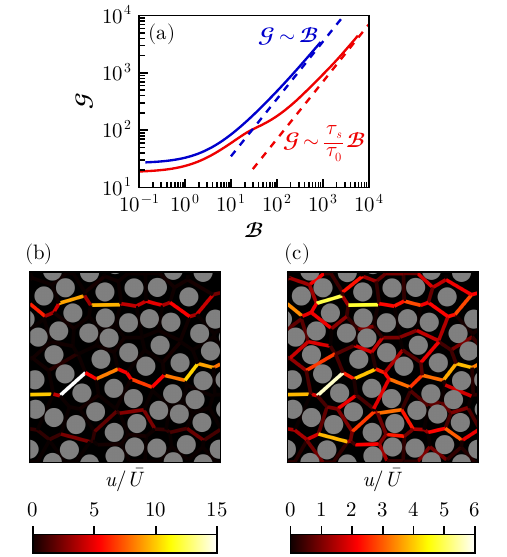}
    \caption{Effect of wall slip on the bulk response and flow topology in geometry G1. (a) Dimensionless bulk pressure gradient $\mathcal{G}$ as a function of bulk Bingham number $\mathcal{B}$ for the no-slip case and for a slipping wall with $\alpha=0.1\,R/K$ and $\tau_s=0.2\,\tau_0$. The dashed lines indicate the large-$\mathcal{B}$ asymptotic scalings, $\mathcal{G}\sim\mathcal{B}$ without slip and $\mathcal{G}\sim(\tau_s/\tau_0)\mathcal{B}$ with slip. (b,c) Velocity maps, normalised by the mean inlet velocity $\bar{U}$, at the same imposed pressure gradient, corresponding to $\mathcal{B}=500$ in the no-slip case: (b) no slip and (c) wall slip. Wall slip lowers the bulk resistance and activates a substantially larger number of flow pathways.}
    \label{fig:pressureDrop_slip}
\end{figure}

We now examine the effect of wall slip on the bulk response and flow topology. Fig.~\ref{fig:pressureDrop_slip}(a) compares the dimensionless pressure gradient $\mathcal{G}$ as a function of the bulk Bingham number $\mathcal{B}$ for the no-slip case and for slip at the walls with $\alpha=0.1\,R/K$ and $\tau_s=0.2\,\tau_0$. As expected, slip reduces the resistance over the full range of $\mathcal{B}$, so that a lower pressure gradient is required to sustain the same bulk Bingham number. It also modifies the large-$\mathcal{B}$ asymptote. In the no-slip limit, the strongly viscoplastic response satisfies $\mathcal{G}\sim \mathcal{B}$, whereas with slip the asymptotic scaling becomes $\mathcal{G}\sim (\tau_s/\tau_0)\,\mathcal{B}$. This reflects the fact that, once the bulk flow is strongly constrained, the dominant threshold is no longer set solely by bulk yielding, but also by the finite wall-slip yield stress.

The effect of slip on the flow architecture is illustrated in Fig.~\ref{fig:pressureDrop_slip}(b,c), which compares the velocity maps at the same imposed pressure gradient, corresponding to $\mathcal{B}=500$ in the no-slip case. Without slip, the flow is confined to only two main percolating pathways. With slip, by contrast, a much larger fraction of the network remains active and several secondary paths reopen. Wall slip therefore weakens the strong channelisation observed in the no-slip case by reducing the local resistance of individual throats and allowing transport through pathways that would otherwise remain blocked. In this sense, slip alters not only the bulk pressure drop (at the same flow rate) but also the connectivity of the flowing network.

To the best of our knowledge, the inclusion of wall slip in a pore-network model for viscoplastic flow through porous media has not been reported previously, despite the fact that slip is common in yield-stress materials and can substantially modify both the threshold for motion and the distribution of flow. The present results show that incorporating slip is essential for predicting both the bulk departure from Darcy behaviour and the topology of the active flow pathways.

\begin{figure*}[ht!]
    \centering
    \includegraphics[scale=1.0]{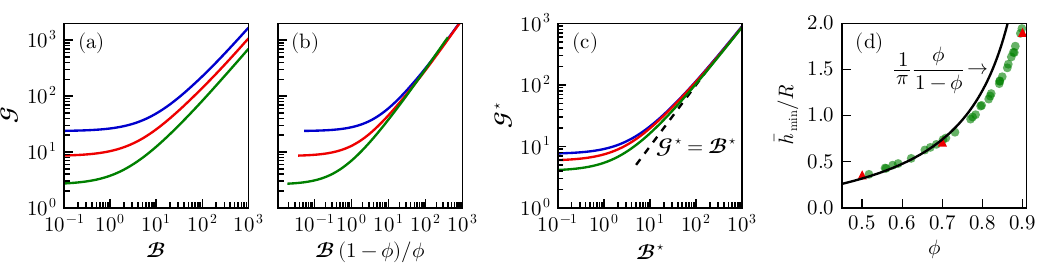}
    \caption{Comparison of scalings for the bulk response of porous-media geometries G1 (blue), G2 (red), and G3 (green). Panels (a-c) show the dimensionless bulk pressure gradient versus bulk Bingham number using (a) the scaling of \citet{Chaparian-Tammisola:2021}, (b) the revised scaling of \citet{Chaparian:2024}, and (c) the present scaling based on $\bar{h}_{\min}$. (d) Dimensionless geometric length scale $h_c/R$ versus porosity $\phi$. Symbols denote random realisations of porous media (green) and the three geometries of \citet{Chaparian-Tammisola:2021} (red). \textcolor{black}{The solid black curve corresponds to the effective length-scale form of the scaling of \citet{Chaparian:2024}, with prefactor $1/\pi$.}}
    \label{fig:scaling}
\end{figure*}

We now examine the near-yield behaviour of the flow. The obstacle-radius scaling of \citet{Chaparian-Tammisola:2021}, although convenient for comparison, does not collapse the $\mathcal{G}$--$\mathcal{B}$ curves across geometries and porosities [Fig.~\ref{fig:scaling}(a)]. In the \textcolor{black}{near-yield,} plastic-dominated regime, flow is strongly channelised \textcolor{black}{and dissipation is concentrated along} the active percolating pathways. \textcolor{black}{This suggests an effective dissipative length scale, $h_c$, associated with the last active percolating pathway near yield, which reflects both the controlling constrictions and, implicitly, the tortuosity of the pathway.} Near yielding, the \textcolor{black}{characteristic} wall stress \textcolor{black}{along this pathway} is $O(\tau_0)$, so
\begin{equation}
  \frac{\Delta p}{L}\sim \frac{\tau_0}{h_c}.
  \label{eq:scaling_percolation_lim}
\end{equation}
Retaining the bulk groups of \citet{Chaparian-Tammisola:2021}, this suggests
\begin{equation}
  \mathcal{G}\sim \frac{\mathcal{B}}{h_c/R}.
  \label{eq:GsimB_over_eps}
\end{equation}

A recent analysis by \citet{Chaparian:2024} proposes a porosity-based normalisation for the onset of percolation,
\begin{equation}
    \frac{\tau_0}{R(\Delta p/L)} \sim \frac{\phi}{1-\phi},
    \label{eq:scaling_chaparian}
\end{equation}
which corresponds to $h_c/R\sim \phi/(1-\phi)$ (with an empirically fitted prefactor). \textcolor{black}{We note that the notation differs from that of \citet{Chaparian:2024}: in the present work $\phi$ denotes the porosity, whereas in \citet{Chaparian:2024} $\phi$ denotes the solid area fraction. Equation~\eqref{eq:scaling_chaparian} has therefore been recast using the present porosity definition.} \textcolor{black}{This effective length scale should not be identified solely with the local width of the critical channel: it also absorbs the effect of the relative channel length, or tortuosity, on the pressure gradient required to sustain flow. The proposed scaling}  substantially improves collapse [see Fig.~\ref{fig:scaling}(b)], but residual offsets across geometry remain.

\begin{figure}[ht!]
    \centering
    \includegraphics[scale=1.0]{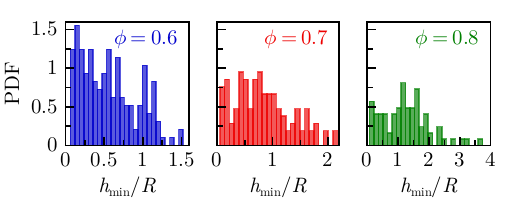}
    \caption{\textcolor{black}{Distributions of the normalised minimum throat half-width, $h_{\mathrm{min}}/R$, for the porous media G1 (blue), G2 (red), and G3 (green). The probability density functions illustrate the variability of constriction widths across each disordered network.}}
    \label{fig:histogram}
\end{figure}

Guided by throat-scale dissipation, we instead tie the dissipative length to a direct constriction statistic and take
$h_c~\sim~\bar{h}_{\min}$, the domain-averaged minimum inter-obstacle half-gap. 
\textcolor{black}{For each throat in the Voronoi network, $h_{\min}$ is calculated from the centre-to-centre distance between the two neighbouring obstacles that define the throat (see Fig.~\ref{fig:geometrynetwork}). The mean value $\bar{h}_{\min}$ and standard deviation $\sigma_{h_{\min}}$ for the whole network reported in Table~\ref{tab:geometric parameters} are then computed over all throats in the network. Figure~\ref{fig:histogram} shows the corresponding probability density functions of $h_{\min}/R$ for G1-G3. Increasing porosity shifts the distributions towards larger throat widths, while their finite spread reflects the disorder of the networks.} Replacing $R$ by $\bar{h}_{\min}$ in the definitions of new groups $\mathcal{B}^\star$ and $\mathcal{G}^\star$ gives
\begin{equation}
\label{eq:star_groups}
\mathcal{B}^\star=\frac{\tau_0}{K(\bar{U}/\bar{h}_{\min})^n},
\qquad
\mathcal{G}^\star=\frac{\Delta p}{L}\,\frac{\bar{h}_{\min}}{K(\bar{U}/\bar{h}_{\min})^n},
\end{equation}
for which \eqref{eq:scaling_percolation_lim} predicts the master behaviour $\mathcal{G}^\star \sim \mathcal{B}^\star$ in the plastic-dominated regime. Fig.~\ref{fig:scaling}(c) shows that, at large $\mathcal{B}^\star$, all branches collapse onto the universal relation $\mathcal{G}^\star~=~\mathcal{B}^\star$, with negligible residual dependence on geometry. The proposed scaling therefore removes the geometry-dependent prefactor present in the original variables and identifies the dissipation-relevant length scale that governs the common linear response in the plastic-dominated regime. \textcolor{black}{It is not intended as a comprehensive Darcy-type law over the full range of Bingham numbers. In the opposite Newtonian limit, dissipation is distributed throughout the pore space, so a single constriction length scale is not expected to collapse the response. A framework that combines the Newtonian and yield limits to describe the pressure-flow response across the full range of $\mathcal{B}$ has recently been proposed by \citet{Chaparian2025}.}

Figure~\ref{fig:scaling}(d) shows that the measured values of $\bar{h}_{\min}$ from multiple random realisations of porous media collapse onto a single porosity-dependent master curve across the geometries considered here. The porosity-based metric of \citet{Chaparian:2024} follows the same overall trend, but departs from the data at high porosity, $\phi \gtrsim 0.7$. This mismatch is likely responsible for the residual offsets left by that scaling, whereas rescaling with the directly measured constriction width $\bar{h}_{\min}$ produces the improved collapse of the bulk pressure gradient in the plastic-dominated regime shown in Fig.~\ref{fig:scaling}(c).

In summary, we have developed a pore-network model for Herschel-Bulkley flow in disordered porous media, including wall slip, that is specified entirely by rheological parameters and pore geometry, without fitting to flow data. Using this framework, we show that near-yield pressure losses are governed by constriction statistics rather than by an obstacle-scale length. Accordingly, rescaling with the domain-averaged minimum throat width collapses the plastic-dominated response across porosities, identifying the dissipation-relevant geometric scale that controls viscoplastic transport in this regime. The collapse is notable because $\bar{h}_{\min}$ is a statistical descriptor of the entire network rather than a direct measure of the final percolating pathway or a strict similarity parameter. In principle, media with the same $\bar{h}_{\min}$ may still differ in connectivity, tortuosity, and other aspects of pore topology. Its success as a scaling length therefore suggests that, across the geometries considered here, $\bar{h}_{\min}$ acts as an effective proxy for the constriction statistics that control the last active pathways near yielding.

The framework also captures the leading influence of wall slip on both bulk resistance and flow topology, demonstrating that slip can substantially reduce the pressure gradient required for transport and activate pathways that remain blocked in the no-slip case. These results show that the breakdown of Darcy behaviour in yield-stress porous-media flow can be captured, and physically interpreted, within a reduced-order description. \textcolor{black}{The reduced-order character of the model is central to its usefulness because it preserves the leading rheological and geometrical controls on the pressure-flow response while avoiding the cost of fully resolved simulations.}

The remaining discrepancies close to yielding are physically informative. The systematic underprediction of resistance is consistent with the fact that the dominant percolating pathway favours the widest constrictions, where the present throat-scale closure is least accurate because of its unidirectional-flow assumption. Moreover, fully resolved simulations \citep{Chaparian-Tammisola:2021} show additional dissipation and occasional branching within pore bodies, which are not captured when pores are represented as lossless nodes. These effects are most important precisely in the near-yield regime, where only a sparse set of pathways remains active. They therefore point to clear next refinements of the reduced-order description, rather than to a limitation of the physical picture itself.

\textcolor{black}{The present formulation is restricted to two-dimensional porous media. Extension to three-dimensional systems would require re-deriving the throat-scale pressure-flow relation for the relevant three-dimensional constriction geometry, and replacing the minimum half-gap $h_{\min}$ by an appropriate descriptor such as a hydraulic aperture or minimum cross-sectional area. The network connectivity, percolation threshold, and pathway tortuosity would also differ in three dimensions. Nevertheless, the mechanisms identified here, namely yielding-induced channelisation, slip-induced reduction of local resistance, and the dominant role of throat constrictions in setting near-yield pressure losses, are expected to remain relevant in three-dimensional systems.} Overall, the present work provides both a predictive reduced-order tool and a physical basis for analysing yielding, channelisation, and slip in complex porous media, with broader relevance to non-Newtonian transport in disordered media.

\emph{Acknowledgements}---The authors thank Qi Chen and James Shemilt for helpful discussions during model development. The authors also acknowledge support from Research IT and use of the Computational Shared Facility at The University of Manchester.

\emph{Funding}---This work was supported by an EPSRC Prosperity Partnership with Unilever through the Centre for Advanced Fluid Engineering and Digital Manufacturing (CAFE4DM), grant EP/R00482X/1. The authors also acknowledge support from the JSPS Overseas Challenge Program for Young Researchers, which funded an exchange placement enabling this collaboration.

\bibliography{aipsamp}

\end{document}